\documentclass[iop,twocolumn]{emulateapj}
\begin{document}

\title{Parallactic Motion for Companion Discovery: An M-Dwarf Orbiting Alcor}

\author{Neil Zimmerman\altaffilmark{1,2},
Ben R. Oppenheimer\altaffilmark{2,1},
Sasha Hinkley\altaffilmark{3,4},
Douglas Brenner\altaffilmark{2},
Ian R. Parry\altaffilmark{5},
Anand Sivaramakrishnan\altaffilmark{2,6,7},
Lynne Hillenbrand\altaffilmark{3}, 
Charles Beichman\altaffilmark{8,9},
Justin R. Crepp\altaffilmark{3},
Gautam Vasisht\altaffilmark{9},
Rick Burruss\altaffilmark{9},
Lewis C. Roberts Jr.\altaffilmark{9},
David L. King\altaffilmark{5},
R\'{e}mi Soummer\altaffilmark{10},
Richard Dekany\altaffilmark{11},
Michael Shao\altaffilmark{9},
Antonin Bouchez\altaffilmark{11},
Jennifer E. Roberts\altaffilmark{9},
Stephanie Hunt\altaffilmark{5}}

\altaffiltext{1}{Department of Astronomy, Columbia University, 550 W 120$^{\mbox{\tiny{th}}}$ St, New York, NY 10027; e-mail address: neil@astro.columbia.edu}
\altaffiltext{2}{Astrophysics Department, American Museum of Natural History, Central Park West at W 79$^{\mbox{\tiny{th}}}$ St, New York, NY 10024; e-mail address: bro@amnh.org}
\altaffiltext{3}{Department of Astronomy, California Institute of Technology, 1200 E. California Blvd., MC 249-17, Pasadena, CA 91125} 
\altaffiltext{4}{Sagan Fellow}
\altaffiltext{5}{Institute of Astronomy, University of Cambridge, Madingley Road, Cambridge CB3 OHA, UK}
\altaffiltext{6}{Department of Physics and Astronomy, Stony Brook University, Stony Brook, NY 11794}
\altaffiltext{7}{NSF Center for Adaptive Optics, Santa Cruz, CA 95064}
\altaffiltext{8}{NASA Exoplanet Science Institute, California Institute of Technology, Pasadena, CA 91125}
\altaffiltext{9}{Jet Propulsion Laboratory, California Institute of Technology, 4800 Oak Grove Dr., Pasadena CA 91109}
\altaffiltext{10}{Space Telescope Science Institute, 3700 San Martin Dr, Baltimore, MD 21218, USA}
\altaffiltext{11}{Caltech Optical Observatories, California Institute of Technology, Pasadena, CA 91125}

\submitted{Submitted to ApJ on 2009 November 11; accepted by ApJ on 2009 December 7}

\begin{abstract}

The A5V star Alcor has an M3-M4 dwarf companion, as evidenced by a novel
astrometric technique. Imaging spectroscopy combined with adaptive optics
coronagraphy allowed for the detection and spectrophotometric characterization
of the point source at a contrast of $\sim$6 $J$- and $H$-band magnitudes and
separation of 1\arcsec~ from the primary star. The use of an astrometric pupil
plane grid allowed us to determine the projected separations between the
companion and the coronagraphically occulted primary star to $\le3$
milliarcsecond precision at two observation epochs. Our measurements
demonstrate common parallactic and proper motion over the course of 103 days,
significantly shorter than the period of time needed for most companion
confirmations through proper motion measurements alone. This common parallax
method is potentially more rigorous than common proper motion, ensuring that
the neighboring bodies lie at the same distance, rather than relying on the
statistical improbability that two objects in close proximity to each other on
the sky move in the same direction. The discovery of a low-mass ($\sim$0.25
M$_\odot$) companion around a bright ($V = 4.0^{\rm m}$), nearby ($d =$ 25 pc)
star highlights a region of binary star parameter space that to date has not
been fully probed.

\end{abstract}


\section{Introduction}

High-contrast imaging is a technique being developed for the study of faint
objects in the vicinity of the closest stars to the Sun, to advance our
understanding of binary stars, substellar companions, exoplanets, and
circumstellar disks. For a recent discussion of this subject, see
\cite{OppHinkley09}. In general, the detection of a point source next to a
bright star is insufficient evidence to establish a physical association. Over
the years, a number of claims of companion detection relying only on single
epoch observations, and a measurement of color have subsequently been disproved
through astrometric measurements. For example, the companion reported in the
\cite{MccarthyProbstLow1985} study of VB 8 was subsequently shown to actually
be a background star~\citep{perrier}. As a result, researchers in this area
have been careful to confirm through astrometry that any putative companion
found shares the proper motion of the primary star, with orbital motion
generally measured after several years of observations.

In fact, most of the stars in surveys for faint companions exhibit appreciable
parallactic motion in addition to their proper motion. For example, over the
course of one year, a star at a distance of 100~pc will appear to trace an
ellipse in the sky with a circumference of roughly 60 mas. The segment of the
curve traversed by this star over an observation baseline of $\sim3$ months
provides an opportunity to discriminate against background stars in the same
manner enabled by common proper motion analysis over longer time
scales~\citep[e.g.,][]{mugrauer}. If the supposed companion maintains the same
offset from the primary star over the duration of time between the observation
epochs---to within an appropriate tolerance set by the upper limit of
hypothetical orbital motion---then a strong argument can be made for the
physical association of the two objects.

We note that the use of parallactic motion discrimination requires higher
precision astrometry than has typically been possible in high contrast imaging.
For example,~\cite{thalmann} achieved a 10 mas level of precision and managed a
detection of common parallax.  Here we achieve a factor of three better
precision to confirm an object's physical association. Other coronagraphs have
not yet demonstrated similar levels of relative astrometry, with tens to
hundreds of milliarcsecond astrometry being typical. This is particularly true
when no other stars with well-established astrometric parameters lie in the
field of view---a common situation that most high-contrast imaging devices face
into the future.

We have used the common parallax method to discover and confirm a companion
orbiting the star Alcor (also known as HD 116842 and HIP 65477; J2000
coordinates $\alpha = 13^{\rm h}25^{\rm m}13.538^{\rm s}, \delta =
+54\degr59\arcmin16.65\arcsec$ in~\cite{hipparcos}).  See the Appendix for a
discussion of Alcor's rich role in the early stage of modern astronomy. While
our astrometry measurements alone permit concrete affirmation of companionship,
we also obtained low-resolution spectra and photometry in the $J$ and $H$
bands, completing the portrait and identifying the companion as an M3-M4 main
sequence star of roughly 0.25 M$_\odot$.  Although Alcor has been surveyed for
possible companions in the past with speckle interferometry, the dynamic range
of this technique at angular separations beyond several times the instrument's
Rayleigh resolution limit is inferior to that obtainable with adaptive optics
coronagraphy, as used in this study. For example, when~\cite{mcalister}
conducted speckle interferometry observations of Alcor with the 3.6 m
Canada-France-Hawaii Telescope, their dynamic range was limited to 3 magnitudes
at separations $>0.04\arcsec$, and consequently could not have detected the
object we describe in this article. On the other hand, Lyot coronagraphs
coupled with adaptive optics systems can routinely attain dynamic ranges of
$\sim 10$ magnitudes at a separation of $1\arcsec$ from the target
star~\citep{OppHinkley09}. Although few low-mass stellar companions to A stars
like Alcor have been imaged, with the increasing prevalence of high contrast imaging
surveys, recently other systems of similar nature have been
found~\citep[e.g.][]{zetavir, kouwenhoven}.

Alcor is a member of the nucleus of the Ursa Major (UMa) moving group. With a
spectral type of A5V, it is one of seven main sequence A stars with high
confidence association to the group, based on kinematic and spectroscopic
indicators~\citep{king}. Despite the long history of studies of the UMa group,
there remains a considerable uncertainty in the age of these stars. After
compiling the photometry of a kinematically selected sample and comparing the
resulting color-magnitude diagram with stellar evolution models,~\cite{king}
arrive at an age estimate of 500$\pm$100 Myr for the group. Another recent
study found that the color-magnitude diagram of the UMa group was best fit with
an isochrone corresponding to an age of 400 Myr~\citep{castellani}. It should
be noted that both of these age estimates are greater than the 300 Myr ages
obtained from earlier work~\citep[e.g.][]{soderblom}. 

For several reasons, Alcor is an attractive target for high contrast imaging
surveys. First, the combination of close distance from the Sun, 24.9 $\pm$ 0.4
pc~\citep{hipparcos}, and its relatively young age (as mentioned above)
increases the probability of detecting a previously unknown substellar
companion: stars closer to the Sun have companions with larger angular
separations on average, and, because substellar objects cool as they age,
younger objects are easier to detect. (See, for example, the cooling
characteristics in~\cite{burrows}.) Furthermore, theoretical models of
fragmentation in circumstellar disks suggest an abundance of low mass
companions around A stars~\citep{kratter}. Indeed, recent direct imaging
discoveries of substellar companions support this
hypothesis~\citep{hr8799,kalas}.

The high apparent brightness of Alcor ($V = 4.0^{\rm m}$) relative to other
nearby stars is yet another agreeable feature. High contrast imaging surveys
rely on the wave front correction provided by adaptive optics (AO) systems to
attain large dynamic ranges within close angular separation of the target star.
When the AO system uses on-axis light rather than an artificial guide star to
measure the wave front errors caused by the atmosphere---as is the case of our
study---the quality of the correction depends strongly on the brightness of the
target star~\citep{troy}. For the above reasons, we chose to include Alcor in
the Project 1640 survey of nearby stars.

\section{Observations}

Project 1640 is a near-infrared, integral field spectrograph situated behind an
Apodized Pupil Lyot Coronagraph (APLC)~\citep{p1640}. During operation, Project
1640 is mounted behind the PALAO adaptive optics system~\citep{dekany} on the
the 200'' Hale Telescope at Palomar. The APLC consists of a pupil plane
apodizer, a hard-edge focal plane mask, and a Lyot stop. The prolate
apodization function and other masks are optimized to deliver broadband
quasi-achromatic starlight suppression~\citep{apodizer,gpitestbed}. The APLC
also includes a fine guidance system and an atmospheric dispersion corrector.
In addition to the apodizer, another novel feature present in the pupil plane
of the APLC is an astrometric grid that serves to indicate precisely the
position of the star when it is occulted by the 370 milliarcsecond diameter
focal plane mask.

The grid of thin opaque lines in the pupil plane produces a periodic linear
array of faint images of the obscured star along the symmetry axes of the grid,
with the star itself at the intersection of two the linear
arrays~\citep{astrometricgrid,maroisgrid}. These satellite spots form an array
of stellar PSFs with angular spacing  $\lambda/d$ ($d$ being the line spacing,
as projected back to the entrance pupil, $\lambda$ the wavelength of the light
forming the image), and brightness approximately $(t/d)^2$ relative to the
central unocculted PSF (where $t$ is the line thickness).  We arranged to have
the closest four satellite PSFs miss the focal plane mask but still lie within
the field of view, to provide stable astrometric fiducials visible in every
coronagraphic image.

Upon exiting the coronagraph, the optical beam passes through an array of 200 x
200 lenslets in the spectrograph. A dispersing prism produces an individual
spectrum corresponding to each lenslet on the 2048 x 2048 pixel infrared
detector, with a spectral resolution of $\lambda/\Delta\lambda\sim30$ between
1.10 $\mu$m and 1.76 $\mu$m ($J$ and $H$ bands). The detector subtends a field of
view approximately $4\arcsec$ in diameter~\citep{p1640}.

Table~\ref{tab:observ} summarizes our observations of Alcor. On 2009 March 16
we obtained 1912 seconds of data with Alcor occulted by the coronagraph, at an
airmass of 1.10, under seeing conditions near $1\arcsec$.  The adaptive optics
system corrected this seeing such that images at 1.65 $\mu$m exhibited a Strehl
ratio of roughly 50\%.  The pupil plane grid used during this observation
produced four astrometric spots in the image at a brightness of $\sim$8
magnitudes fainter than the target star. A point source was immediately
noticeable $\sim1\arcsec$ from Alcor in the raw data. We observed Alcor again
with good atmospheric conditions on 2009 June 27, this time obtaining a total
of 293 seconds of occulted data.  Again the point source of interest was visible, in
roughly the same location with respect to Alcor.  During the June observations
we used a pupil grid with thicker reticule wire, providing brighter astrometric
spots, $\sim$6 magnitudes fainter than Alcor. 

\begin{deluxetable*}{cccccc}
\tabletypesize{\footnotesize}
\tablecaption{Summary of Project 1640 observations of the Alcor System.\label{tab:observ}}
\tablehead{\colhead{Mean UT Date} & \colhead{Besselian Year} & \colhead{$\tau_{\mbox{exp}}$ (s)} & \colhead{$\lambda$ ($\mu$m)} & \colhead{$\rho$ (mas)} & \colhead{P. A. (Degrees East of North)}}
\startdata
2009 March 16 10:35 & 2009.20469 & 1912 & 1.10-1.76 & 1050 $\pm$ 1 & 206.5 $\pm$ 0.1 \\
2009 June 27 3:51 & 2009.48593 &  293 & 1.10-1.76 & 1043 $\pm$ 1 & 207.1 $\pm$ 0.1 \\
\enddata
\end{deluxetable*}

\section{Data Processing}

The Project 1640 integral field spectrograph (IFS) produces information with
three dimensions simultaneously: two spatial and one spectral. Therefore, the
most natural way to view the data is in the form of a cube where each slice is
an image of the target field in a particular wavelength channel. We devised a
data pipeline to automate the process of converting the detector images---each
containing a mosaic of $4\times10^4$ closely packed spectra---into a data cube.
The complete description of the details of this technique will be published
elsewhere.  Here we provide a general overview of how it works.

An essential component of the cube extraction is a library of images made by
illuminating the IFS with a tunable laser. Each of these laser images contains
the response of the IFS to a specific wavelength: a matrix of illuminated spots
corresponding to the individual lenslets of the IFS.  Effectively, they are
keys showing what regions of the $4\times10^4$ spectra landing on the detector
correspond to a particular central wavelength. The data pipeline uses the laser
images to extract the science data and map them onto a cube, forming
twenty-three images at wavelengths between 1100 and 1760 nm, each with a
bandwidth of 30 nm. In addition to the mapping between the detector plane and
the data cube, the Project 1640 data pipeline carries out numerous steps to
prepare the data for analysis, including bias/dark-subtraction, bad pixel
correction, and flat-fielding.

\begin{figure*}[htb]
\epsscale{0.7}
\plotone{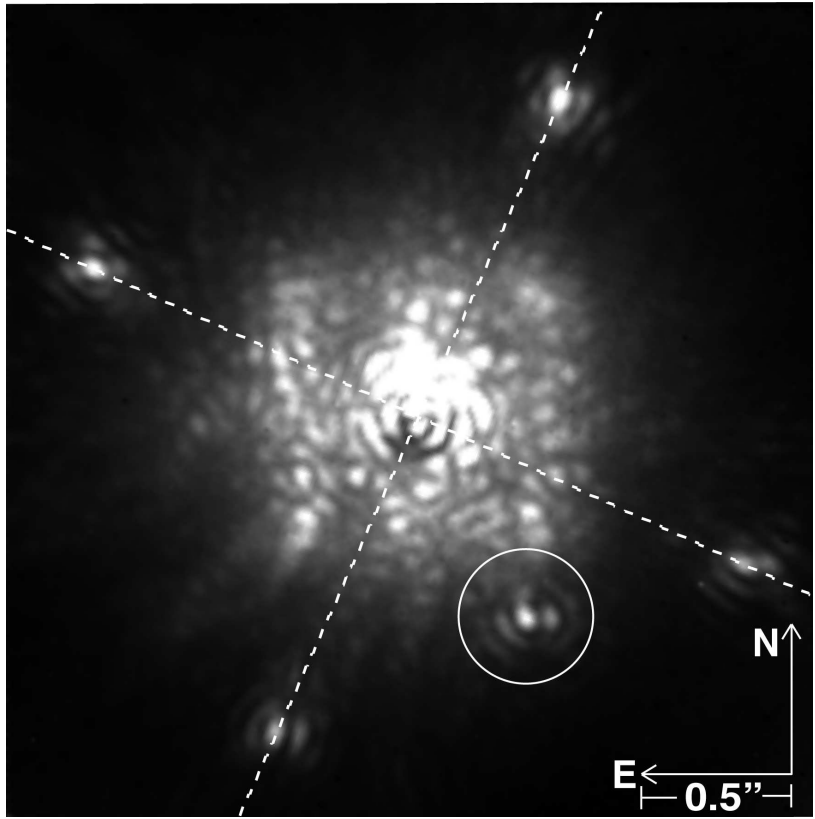}

\caption{Coronagraphic image of Alcor obtained in June of 2009.  This the slice
of the data cube corresponding to central wavelength $\lambda$ = 1.61 $\mu$m.
The dashed lines illustrate the intersection of the lines between the
astrometric spots, indicating the position of Alcor behind the occulting mask.
Coincidentally, the astrometric spots are approximately 6 magnitudes fainter
than the occulted star, similar to the brightness of the companion. The
companion is the circled point source south-west of the image
center.\label{fig:intersect}}

\end{figure*}

Figure~\ref{fig:intersect} shows the $1.61 \mu$m slice of a data cube formed
from our 2009 June 27 data. It is the result of 40 detector reads, each of
duration 7.7 seconds, for a total integration time of 293 seconds. The four
astrometric spots are visible on the peripheral of the image, while the point
of source interest is detected south-west of the occulting mask. We aligned and
co-added all the data from each epoch, producing one final data cube
representing each epoch.

In the rest of the article we refer explicitly to ``lenslet pixels'' to
describe the pixels comprising the data cube, to avoid ambiguity with the
pixels on the detector of the IFS. Since each lenslet pixel constitutes a
measurement of flux from an area element of the sky within a certain wavelength
range, it can be treated in the same manner as the pixel of an ordinary digital
image. Our analysis is done strictly on the data cubes that have already been
extracted from the detector images, so the detector pixels are absent from
further discussion.

\section{Photometry}

In each spectral channel of the 2009 March 16 data cube, we performed aperture
photometry on the putative companion. The residual light from the primary star
significantly contaminated our image of the source. One component of this noise
is in the form of speckles, which are not distributed in a smooth, easily
modeled fashion~\citep{racine,highstrehl,aime,dynamicrange,soummerspeckles}.
Numerous efforts are underway to develop algorithms that remove speckles from
integral field spectrograph data by exploiting their chromatic
properties~\citep[e.g.][]{sparksandford}. However, these speckle suppression
techniques have not yet matured enough to apply to data from an instrument such
as ours without also altering the measured flux of true point sources. To
minimize the effect of residual light from the primary star on our
measurements, we counted the signal only in the core of the point spread
function, even though up to two Airy rings of the source diffraction pattern
are apparent in the data (as in Figure~\ref{fig:intersect}). Since the point
spread function scales with wavelength, we used a different photometric
aperture size for each half of the operating band to match the core size. For
the first 11 channels (central wavelengths 1.10 $\mu$m-1.40 $\mu$m), we
measured the flux in a circle of radius 3 lenslet pixels, and used a 4 lenslet
pixel radius circle for channels 12-23 (central wavelengths 1.43 $\mu$m-1.76
$\mu$m).

In each channel of the data cube, the contaminating light from the primary star
contributed $\sim$40-50\% of the flux counts within the core photometric
aperture. To account for this, we subtracted a ``background'' estimate formed
from the median of pixel values in the annulus between 16 and 19 lenslet pixels
from the center of the source. The 16 lenslet pixel inner radius of this
background annulus is outside the detected diffraction pattern of the point
source of interest.

The uncertainty in the assumed level of contaminating light from Alcor based on
the annulus median is the dominant source of error in the photometry. We
estimated the uncertainty in the assumed contamination by measuring the scatter
in the median values of carefully chosen patches of the channel images. These
patches were at nearly the same lenslet pixel separation from Alcor as the
putative companion PSF, contained within the 16-19 pixel ``background'' annulus
(so that they were beyond the influence of the putative companion PSF), and had
the same area as our core photometric aperture. In other words, we based our
uncertainty in the subtraction of the primary star's contribution by examining
the behavior of its residual light in parts of the image that are subject to
similar contamination to the core photometric aperture. We find the resulting
error remains $\sim$5\% of the companion signal across the band.

We derived $J$- and $H$-band fluxes of the putative companion using a
reference star observation to calibrate the photometry. On 2009 March 14, two
days before our first epoch of Alcor data, we obtained a 7 second unocculted
exposure of HD 107146/HIP 60074 (apparent magnitude $V$ = 7.04, spectral type
G2V) at an airmass of 1.05 under similar observing conditions. Even though HD
107146 has a known debris disk, it is optically thin and only detected near our
instrument's wavelengths in Hubble Space Telescope data~\citep{ardila}. The HST
data show scattered light distributed in a ring of inner radius $3\arcsec$,
which is outside our field of view and far beyond our $\sim0.1\arcsec$
photometric aperture. 

We carried out aperture photometry on the point spread function in the HD
107146 data cube in an identical fashion as for the source in the Alcor image,
using the same aperture and background annulus sizes. In the 2 Micron All Sky
Survey (2MASS) Point Source Catalog~\citep{2mass}, HD 107146 has $J$- and $H$-
band photometry listed as 5.87$\pm$0.02 and 5.61$\pm$0.02, respectively. We
summed the core fluxes of the point source of interest and HD 107146 in the
channel ranges corresponding to the 2MASS $J$ and $H$ filters (1.13 $\mu$m-1.34
$\mu$m, and 1.46 $\mu$m-1.73 $\mu$m, respectively). Subtracting the raw
magnitudes of the HD 107146 $J$ and $H$ sums from the 2MASS magnitudes, we
derived correction magnitudes for each filter. Applying those corrections to
the channel sums of the putative companion, we arrived at the broadband fluxes
listed in Table~\ref{tab:alcorBphotom}.

\begin{deluxetable}{ccc}
\tablewidth{0pt}
\tablecaption{Near-infrared photometry of Alcor B.\label{tab:alcorBphotom}} 
\tablehead{\colhead{Band} & \colhead{m} & \colhead{M}}
\startdata
$J$ & 9.95 $\pm$ 0.06 & 7.97 $\pm$ 0.06 \\
$H$ & 9.56 $\pm$ 0.06  & 7.58 $\pm$ 0.06 \\
\enddata
\end{deluxetable}

We estimated the probability of a star with matching photometric properties
unassociated with Alcor coinciding with our field of view. One way to do this
is to determine the surface density of point sources that have fluxes within
the two-sided 5$\sigma$ confidence interval of our $J$- and $H$-band
magnitudes, corresponding to flux bounds 25$\%$ above and below our stated
measurements. We queried the 2MASS Point Source Catalog for the number of $J$
and $H$-band point sources in the $2^{\circ}\times2^{\circ}$ area centered on
Alcor's coordinates, separated into one magnitude-wide bins extending between
magnitudes 8 and 16. A linear regression fit to the logarithm of the source
count as a function of magnitude yields the relations $\log_{10}(\mbox{$J$-band
sources deg$^{-2}$}) = -1.764 + 0.284m_{J}$ with a r.m.s. residual of 0.167,
and $\log_{10}(\mbox{$H$-band sources deg$^{-2}$}) = -1.604 + 0.284m_{H}$ with
a r.m.s. residual of 0.090. Integrating these point source surface density
relations between the 5$\sigma$ flux boundaries of the supposed companion,
$9.71 \le m_J \le 10.26$ and $9.32 \le m_H \le 9.87$, we arrive at $J$-band and
$H$-band point source surface densities 3.9 deg$^{-2}$ and 4.2 deg$^2$,
respectively. Taking the larger of these two surface densities, 4.2 deg$^{-2}$,
and multiplying by our $4\arcsec\times4\arcsec$ field of view, we expect
$5.2\times10^{-6}$ sources matching the photometric characteristics of the
supposed companion in a given $4\arcsec\times4\arcsec$ field of view in this
part of the sky. Multiplying this by 100 to roughly account for the number of
stars we have surveyed so far with null detections of stellar companions, we
arrive at a posteriori probability of 0.05\% that the source is unassociated
with Alcor. Later in this article we will demonstrate how our astrometry
reduces this probability to an even less significant quantity. With that
knowledge in hand, we hereafter refer to the point source of interest as Alcor
B, following the traditional nomenclature of directly imaged companions.

The parallax distance modulus of Alcor is 1.98$^m$, so to place Alcor B at the
same distance implies it has absolute magnitudes $M_J = 7.97\pm0.06$ and $M_H =
7.58\pm0.06$. \cite{henry} derived empirical mass-luminosity relationships for
stars with masses between $0.18$M$_\odot$ and $0.50$M$_\odot$. When we we apply
these to our absolute $J$- and $H$- band magnitudes, and take into account the
variance inherent to the model and our own photometric uncertainty, we
calculate mass estimates of 0.26$\pm_{0.07}^{0.10}$ M$_{\odot}$ and
0.21$\pm_{0.03}^{0.04}$ M$_{\odot}$, from the $J$- and $H$-band luminosities,
respectively. When we compare our fluxes to theoretical mass-luminosity models
computed specifically for 600 Myr-old stars by~\cite{baraffemodel}, similar to
the published age estimates of Alcor, we find masses of 0.27M$_\odot$ and
0.25M$_\odot$, respectively. According to the mass-spectral class relationship
for low mass stars derived by~\cite{baraffe}, a star with a mass between
0.2M$_\odot$ and 0.3M$_\odot$ indicates a spectral type in the range from M2V
to M3.5V.

\section{Spectroscopy}

We extracted a low-resolution spectrum of the stellar companion from the IFS
data. Again, we used the star HD 107146 as a reference source. As stated above,
even though HD 107146 has a disk, it is faint and outside our field of view.
Furthermore, the star lacks the excess emission that some disk hosts possess at
10 $\mu$m~\citep{metchevIR}, so to the best of our knowledge, the spectrum is
ordinary for a star of its class in our wavelength regime.

We began the spectral calibration by determining channel-wise corrections for
the wavelength-dependent transmission of the atmosphere and instrument. To do
this, we made the assumption that at the spectral resolution of our data cube
($\lambda / \Delta \lambda \sim 30)$, and within our photometric errors, the
spectrum of HD 107146 matches that of a typical G2V star. We compared our raw
spectrum of HD 107146 with the measurement by~\citep{rayner} of the
near-infrared spectrum of HD 76151, another G2V star. The HD 76161 data is part
of a suite of reference stellar spectra collected at NASA's Infrared Telescope
Facility (IRTF). We binned the publicly archived HD 76151 spectrum to our
spectrograph's resolution, divided it into our raw HD 107146 spectrum, and
mean-normalized the result to obtain our response vector. 

We obtained our raw spectrum of Alcor B by carrying out aperture photometry on
the reduced data cube in the same manner as described in the Photometry
section: counting the signal in an aperture containing the core of the PSF and
subtracting the median of an annulus around the source multiplied by the
aperture area. As before, our photometric errors were dominated by the
uncertainty in the annulus estimate of Alcor's residual light in the companion
photometric aperture, which includes a smooth halo and a speckle component.

\begin{figure*}
\epsscale{0.7}
\plotone{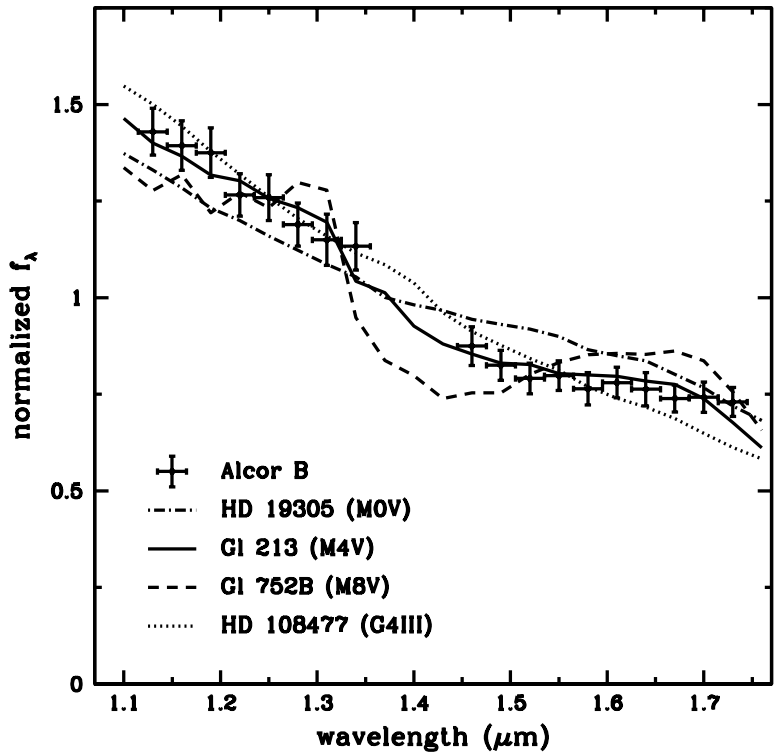}
\caption{Spectrum of Alcor B extracted from the 2009 March data, compared with three examples of M-dwarf spectra and a giant spectrum whose $J-H$ color matches the photometry of the companion.\label{fig:alcorBspectrum}}
\end{figure*}

We divided our raw Alcor B spectrum by the response vector to obtain the
calibrated spectrum of the companion plotted in
Figure~\ref{fig:alcorBspectrum}. In our plot we omit the five channels of our
spectral range that are strongly subjected to variable telluric absorption,
those with central wavelengths 1.10 $\mu$m, 1.37 $\mu$m-1.43 $\mu$m, and 1.76
$\mu$m. The spectrum data points are normalized to the mean of the included
channels.

We compared the spectrum of Alcor B with a broad range of examples of M-dwarf
spectra from the IRTF Spectral Library~\citep{rayner}. In addition, we compared
our companion spectrum with that of the giant star in the IRTF Spectral Library
with the closest $J-H$ color, HD 108477, a G4III star with a 2MASS $J-H$ color
of 0.34. After rebinning all of these comparison spectra to our data cube's
spectral resolution, we normalized them and calculated the root mean square
differences from the Alcor B spectrum. The spectra of three of these reference
spectra are plotted alongside the Alcor B data in
Figure~\ref{fig:alcorBspectrum}. Of all of the spectra compared, the two
closest matches to Alcor B are the M3.5V star Gl 273 and the M4V star Gl 213,
both with root mean square residuals of 4\%.  Although the shape of the G4III
giant spectrum is qualitatively different from the Alcor B data, particularly
in terms of the slope across $H$ band, its fit has a r.m.s. residual of only
5\%.  This serves to indicate that at this spectral resolution there is
ambiguity in discriminating between G-giant and M-dwarf stars of similar color. 

\section{Astrometry}

\par 

In each channel of the data cube, the intersection of the two perpendicular
lines formed by the four astrometric grid spots determines the location of
Alcor on the lenslet array (Figure~\ref{fig:intersect}). We compared these
locations with the directly measured position of the companion in the data to
measure the relative offset at each epoch. Due to a slight residual atmospheric
dispersion causing an apparent drift in the position of the star by $\sim$2
lenlsets over the wavelength range in the data cube, we only compared the spot
intersection with the position of the companion measured within the same
channel. The companion and the grid spots have strongest detections in a subset
of cube channels in the $H$ band, enabling the most accurate position
determination at those wavelengths. These are also the wavelengths at which the
atmospheric dispersion corrector is optimized. In addition, as is generally the
case, the wave front correction of the AO system is better at $H$ band than in
$J$ band. Therefore, we used only the five channels from 1.55 $\mu$m to 1.67
$\mu$m---those spanning the H-band transmission peak---to deduce the relative
offsets at each epoch. We measured the positions of the companion and the grid
spots in the data cubes by fitting Gaussian profiles to the point spread
functions. For each epoch, we took the mean of the offsets between the grid
spot intersection and the companion PSF in the five aforementioned channels to
arrive at our final estimates. We then applied the Student's t distribution (as
appropriate when estimating a mean from a sample of five
measurements---see~\cite{dean}) to derive 68\% confidence intervals based on
the standard deviation of the offset components between the channels. 

To convert the lenslet pixel offsets into angular offsets oriented with
equatorial coordinates, we applied our plate scale of $19.2\pm0.1$ mas/lenslet,
and compensated for the rotation of our detector (the columns of the extracted
data cubes are oriented $70.6\pm0.1 ^\circ$ counter-clockwise with respect to
north). Both the plate scale and rotation were derived from a series of
observations of calibration binary systems with Grade 1 orbit solutions in the
USNO Sixth Orbit Catalog~\citep{orbitcat} between July 2008 and March 2009.
Standard errors were propagated through all calculations to reflect 68\%
confidence intervals in the error bars. In Table~\ref{tab:astrometry} we list
the resulting offset components between Alcor B and its host star. 

As described in the Photometry section, if we consider only our photometric
measurements of the putative companion, we are left with a small possibility
($\sim0.05$\%) that it is an unassociated star coinciding with our line of
sight.  Now, with our astrometry, we can rule out this possibility to a
stronger degree, in order to affirm the physical association with Alcor.

\begin{figure*}
\epsscale{0.95}
\plotone{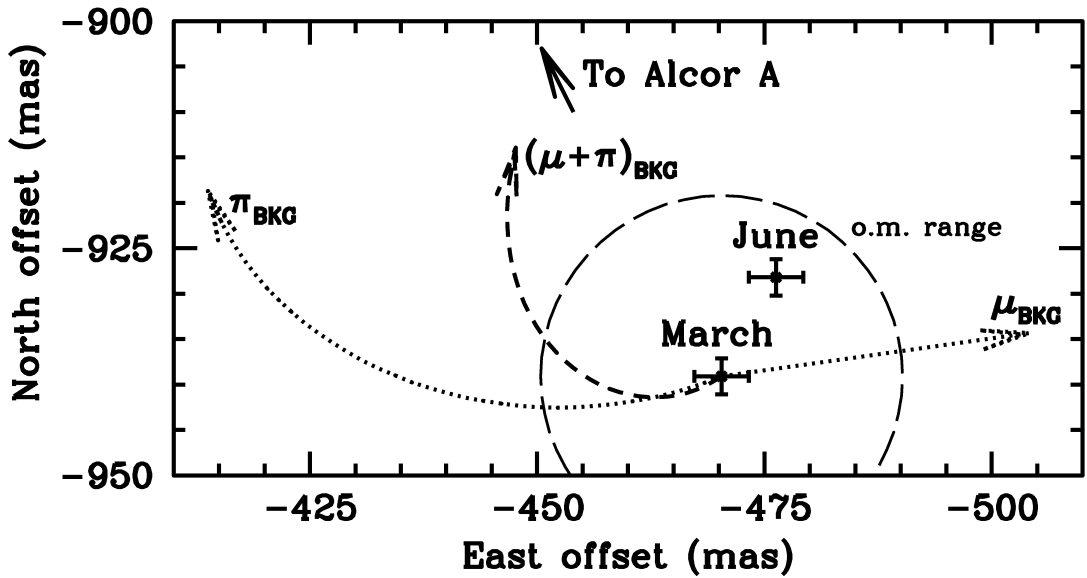}

\caption{Summary of astrometry measurements, plotted in terms of north and east
offsets from Alcor. The two position measurements of Alcor B are plotted with
their associated 1-sigma error bars, and labeled March and June, corresponding
to UT epochs 2009 March 16 and 2009 June 29, respectively. The
$\mu_{\mbox{\tiny{BKG}}}$ and $\pi_{\mbox{\tiny{BKG}}}$ arcs shows the expected
change in offset of a fixed background star due to Alcor's proper and
parallactic motion. The $(\mu + \pi)_{\mbox{\tiny{BKG}}}$ arc is the resultant
of these components over the course of our two observation epochs. We have also
plotted a circle labeled ``o.m.  range'' containing an estimate of the upper
limit of orbital motion with respect to the March
position.\label{fig:astrometry}}

\end{figure*}

First, as illustrated in Figure~\ref{fig:astrometry}, we can rule out the
simple notion that the supposed companion is actually a distant background star
lacking significant proper or parallactic motion---one that is, for our
purposes, fixed on the sky. For example, one could imagine a luminous star at a
distance of $\sim$1 kpc, whose parallactic motion between our two observations
is only $\sim$1.5 mas, and whose projected space motion also happens to be near
or below our astrometric precision. By contrast, over our 103-day baseline, the
parallactic and proper components of Alcor's motion (see
Table~\ref{tab:alcormotion}) resulted in a displacement with a magnitude of 34
mas. Because our two images remained centered on Alcor over the course of its
motion, a fixed background star in our data would appear to shift about 34 mas
relative to Alcor. More specifically, since the overall apparent motion of
Alcor between our observations was 22.7 mas west and 25.2 mas south, a fixed
background star lying south-west of Alcor would have shifted 22.7 mas east
relative to Alcor (decreasing the magnitude of its offset from Alcor in Right
Ascension), and 25.2 mas north relative to Alcor (decreasing the magnitude of
its offset from Alcor in Declination). The arc labeled $(\mu + \pi)_{BKG}$ in
Figure~\ref{fig:astrometry} represents this circumstance.  Instead, we observed
a westward motion of $6.0\pm4.3$ mas relative to Alcor and a relative northward
motion of only $10.9\pm2.3$ mas (the two positions are labeled ``March'' and
``June'' in Figure~\ref{fig:astrometry}). So the observed motion is
inconsistent with a background star exhibiting a low apparent motion on the
order of several milliarcseconds or less.

Now we consider the case of a distant background star that does exhibit
significant apparent motion, in a such a way that matches the observed
displacement of the companion star between our observation epochs. The least
luminous giant with consistent $J-H$ color, a star of type G2III, would have to
be at a distance of about 740 pc in order for its apparent magnitude to be
consistent with our photometry. By combining our measurement of the change in
Alcor B's offset from Alcor A and our knowledge of the apparent motion of the
primary star, we can deduce the absolute motion of the putative companion on
the sky, decoupled from Alcor: $28.6\pm4.3$ mas west and $-14.4\pm2.3$ mas
south. At a distance of 740 pc, the expected parallactic motion between our
observation epochs is 1.9 mas west and 0.7 mas south. Then, a proper motion of
26.7 mas west and 13.7 south is needed to make up for the difference from the
observed apparent motion. We compute the space velocity from this assumed
proper motion and distance using the formulas described in~\cite{coordxform}.
Assuming zero radial velocity, this star would need a galactic space velocity
of U = -150 km s$^{-1}$, V = -300 km s$^{-1}$, and W = 130 km s$^{-1}$ to be
consistent with the apparent motion we measure.  The largest component of this
space velocity, V, indicates a strong retrograde galactic orbit. For more
luminous giant stars, the necessary space velocities grow to even more unlikely
values---a K1III giant, for example, would have need a V component of -600 km
s$^{-1}$ to be consistent with our astrometry. In that case, V is within the
range of estimates of the local escape speed of the galaxy (e.g., 498 km
s$^{-1} < v_{\rm esc} < 608$ km s$^{-1}$ from ~\cite{smith} and 489 km s$^{-1}
< v_{\rm esc} < 730$ km s$^{-1}$ from~\cite{kochanek}). 

The only plausible scenario remaining, that we have in fact discovered a
low-mass companion to Alcor, can be checked by comparing the measured relative
motion to Alcor with an estimate of the upper limit on the orbital motion a
true companion would exhibit between our two observation epochs. The empirical
mass-luminosity relation for intermediate-mass stars of~\cite{malkov} implies a
mass of 1.8M$_\odot$ for the A5V primary star, given its absolute magnitude
$M_V = 2.01$. Assuming a mass of 0.25 M$_\odot$ for the companion, a circular
orbit of the projected radius 26 AU ($1.05\arcsec$ at 24.9 pc) would have a
period of roughly 93 years, resulting in an apparent motion of $\sim$20 mas if
it were orbiting face-on. In fact, the motion we detected is smaller than this,
but any inclination, eccentricity, or different semi-major axis in the orbit
could change the expected orbital motion. However, most importantly, the
apparent motion of Alcor B that we do detect is consistent with plausible
orbital motion around Alcor. A circle illustrating the range of possible
orbital motion with respect to the position of Alcor B at the first observation
epoch is shown in Figure~\ref{fig:astrometry}.

\begin{deluxetable}{cccc}
\tablewidth{0pt}
\tablecaption{Relative Astrometry of Alcor B\label{tab:astrometry}}
\tablehead{\colhead{Component} & \colhead{2009 March 16} & \colhead{2009 June 27} & \colhead{Change}}
\startdata
East offset (mas) & $-470.3 \pm 3.1$ & $-476.3 \pm 2.9$ & -6.0 $\pm$ 4.3 \\
North offset (mas) & $-939.1 \pm 1.7$ & $-928.2 \pm 1.5$ & 10.9 $\pm$ 2.3 \\
\enddata

\tablenotetext{a}{The equatorial coordinate offsets of Alcor B relative to its
host star on 2009 March 16 and 2009 June 27 (Besselian dates 2009.2047 and
2009.4859, respectively) followed by the change between the two epochs.} 

\end{deluxetable}

\begin{deluxetable}{cl}
\tablewidth{2.5in}
\tablecaption{Apparent Motion of Alcor Between Observation Epochs\label{tab:alcormotion}}
\tablehead{Component & Change (mas)}
\startdata
$\Delta$East$_{\rm PM}$ & 33.9 \\
$\Delta$North$_{\rm PM}$ & -4.8 \\
$\Delta$East$_{\pi}$  & -56.5 \\
$\Delta$North$_{\pi}$  & -20.4 \\
$\Delta$East$_{\rm{PM} + \pi}$ & -22.7 \\
$\Delta$North$_{\rm{PM} + \pi}$ & -25.2 \\
\enddata

\tablenotetext{a}{Subscripts PM and $\pi$ indicate the expected proper and
parallactic motion, respectively. The subscript PM + $\pi$ indicates the total
displacement due to combined proper and parallactic motion.} 

\tablenotetext{b}{Based on values from the Hipparcos Catalog~\citep{hipparcos},
applied to epochs 2009 March 16 and 2009 June 27 (Besselian dates 2009.2047 and
2009.4859, respectively). The Hipparcos tables identify a proper motion of
120.35 mas/yr in R.A. (corrected for Declination to reflect motion on a great
circle), -16.94 mas/yr in Dec (with a 1-$\sigma$ error of $<$0.52 mas/yr in
each direction), and a parallax of 40.19$\pm$0.57 mas.}

\end{deluxetable}

\section{Discussion and Conclusions}

Although we observed Alcor only twice over a baseline of 103 days, the high
precision ($\le$ 3 mas) relative astrometry enabled by the pupil plane grid of
Project 1640 allowed us to find common parallactic and proper motion, thereby
ruling out the possibility that the newly detected point source is a background
star. We expect that as we improve our techniques for interpreting data from
the integral field spectrograph, we can attain yet higher astrometric precision
in future studies. With sufficient sensitivity, such methods can be extended to
lower mass objects, to characterize young, long period exoplanets. The rapidity
of common parallax discrimination, as opposed to observation baselines
$\gtrsim1$ year relying on proper motion analysis alone, could improve the
efficiency of future high contrast imaging efforts. In particular, in the
surveys that will be carried out with Project 1640 and the planned high-order
adaptive optics system for Palomar, as well as the similar system planned for
Gemini Observatory (Gemini Planet Imager; \cite{bruceMacintoshSPIE}), repeated
observations of a faint companion candidates should be scheduled $\sim$1-4
months from the initial detection epoch. This period is short enough for a
typical target to remain visible in the night sky, but long enough to allow for
sufficient parallactic motion for stars closer than $\sim50$ pc.

Under the most favorable observation arrangements, where investigators can
acquire high precision relative astrometry of a possible companion three or
more times within several months of the discovery date, the common parallax
technique can demonstrate physical association with yet greater rigor than we
have achieved here. If the primary star traces a parallactic arc of sufficient
curvature over the observation baseline, three epochs of data indicating a
persistent offset vector can no longer be accounted for geometrically by a
background star. In such a case, the celestial coordinate trajectory of a true
companion would be seen to deviate from regular linear motion to an extent that
cannot be explained by a masquerading background star, even one with the most
anomalous space velocity. To thereby show that the discovered neighbor follows
the arc of the host star's parallactic ellipse would demonstrate companionship
most conclusively.

We note that recently~\cite{thalmann} also used common parallax measurements to
confirm the existence of a companion to the star GJ 758. However, Thalmann et
al. do not discuss the significance of this method in their article. Presumably
since the coronagraph they used lacks an astrometric grid, they were not able
to attain as high a precision in the relative position of the companion as
achieved in our study, reporting an uncertainty of 9.5 mas.

We acquired a low resolution ($\lambda / \Delta \lambda \sim 30$) spectrum of
the companion with the Project 1640 Integral Field Spectrograph, enabling a
preliminary spectral classification of M3V-M4V.  We demonstrated that even with
significant contamination of host starlight, a low spectral resolution integral
field spectrograph can be effective in constraining the spectral type of newly
discovered companions. A comparison between our broadband $J$- and $H$-band
fluxes with two different mass-luminosity relationships yielded mass estimates
ranging from 0.21-0.27 M$_\odot$. Unlike lower mass
stars~\citep[e.g.][]{metchev}, few systematic surveys have been carried out
with AO-equipped telescopes to characterize the frequency and mass ratio
distribution of binary A stars, so it is difficult to place the significance of
this discovery in the context of established binary star properties.

The object we found is relevant to the conundrum of x-ray emission from A
stars.  Unlike lower mass (F-M) main sequence stars and O and B stars, there is
no consensus on a physical mechanism for x-ray emission from A stars. They lack
the energetic winds of more massive stars, which explain the commonly seen
x-ray activity of O and B stars. They also lack the convection-driven magnetic
dynamos of lower mass main sequence stars, which are widely held to be the
source of their x-ray emission~\citep{pallavicini}.  Despite this, 10-15\% of A
stars were detected as x-ray sources by the R{\"o}ntgen Satellite
(ROSAT)~\citep{schroeder}. In fact, Alcor is one of them, detected in the ROSAT
All Sky Survey, with an x-ray luminosity of $L_{X} = 2.8\times10^{28}$ erg
s$^{-1}$. It has long been proposed that unseen lower mass companions could
account for the anomalous x-ray emission of many of these A
stars~\citep{schmitt}. When~\cite{patience} surveyed A stars for stellar
companions with the U.S. Air Force Advanced Electro-Optical System (AEOS), they
found previously unknown companions to 8 of the 11 observed A stars with known
x-ray source coincidence. Our finding lends further support to the hypothesis
that hidden stellar companions explain the majority of perceived A star x-ray
activity. 

\acknowledgments

The authors wish to express our appreciation toward Jean Mueller, Kajsa Peffer,
Karl Dunscombe, and the Mountain Crew at Palomar Observatory. We thank Adam
Burgasser and Emily Rice for sharing comparison spectra to aid in our
classification of the companion star. We are also indebted to our anonymous
referee for expediting our review process. Project 1640 is funded by National
Science Foundation Grants AST-0520822, AST-0804417, and AST-0908484. A portion
of the research in this paper was carried out at the Jet Propulsion Laboratory,
California Institute of Technology, under a contract with the National
Aeronautics and Space Administration (NASA) and was funded by the internal
Research and Technology Development funds. In addition, part of this work was
performed under a contract with the California Institute of Technology
(Caltech) funded by NASA through the Sagan Fellowship Program.  The members of
the Project 1640 team are also grateful for support from the Cordelia
Corporation, Hilary and Ethel Lipsitz, the Vincent Astor Fund, Judy Vale,
Andrew Goodwin, and an anonymous donor.

{\it Facilities:} \facility{Hale (PALAO, Project 1640)}

\newpage

\appendix
\section{History of Early Parallax Measurement Attempts with Alcor}

In celestial lore, Alcor is best known for the place it shares in the sky with
Mizar in the handle of the ``Big Dipper'' asterism. Alcor and Mizar were
commonly used in ancient times as a test of visual acuity~\citep{bohigian}. The
fainter Alcor, at a separation of 12', cannot be discerned unless one has good
natural eyesight or corrective glasses. The pair is collectively designated
$\zeta$ Ursa Majoris in Johann Bayer's 1603 Uranometria star catalog.  Although
Alcor and Mizar share physical association in the Ursa Major moving group, it
has yet to be shown conclusively whether or not they are gravitationally bound.
However, Mizar itself was the first true multiple star system to be resolved
with a telescope, by Benedetto Castelli, a colleague of Galileo
Galilei~\citep{fedele}. 

Alcor and Mizar were also among the subjects of the first attempts to measure
stellar parallax. Well before the invention of the telescope, stellar parallax
was identified as the most conclusive way to demonstrate the Copernican
assertion that the Earth orbits the Sun~\citep{siebert}. In 1597 Johannes
Kepler wrote a letter to Galileo encouraging him to attempt stellar parallax
measurements, hoping he would succeed where Tycho Brahe had failed.  Galileo
recognized that the field of view containing Mizar, Alcor and Sidus Ludoviciana
(also known as HD 116798) was ideal for parallax measurements. The three stars
form an approximate right triangle and they are at high declination, meaning
that parallactic motion would trace an ellipse of low eccentricity. The
triangle provides a position reference in two spatial directions. Galileo spent
considerable effort trying to measure an actual parallactic
motion~\citep{galileo} and distances to stars, but never succeeded. See
Siebert's (2005) article for more detail. Although these early attempts at
measuring parallax were beyond the measurement precision of the time, it is
somehow poetic that a new result concerning Alcor some 400 years later relies
on parallax.

\end{document}